\documentclass[12pt,preprint]{aastex}

\input{epsf}
\journalid{}{}
\articleid{}{}

\shorttitle{Fast X-ray Transients and Their Connection to GRBs}
\shortauthors{Arefiev, Priedhorsky, \& Borozdin}

\begin{document}
\def\gtorder{\mathrel{\raise.3ex\hbox{$>$}\mkern-14mu
             \lower0.6ex\hbox{$\sim$}}}
%

\title{ Fast X-ray Transients and Their Connection to Gamma-Ray Bursts}

\author{Vadim A. Arefiev\altaffilmark{1}, William C. Priedhorsky, \& Konstantin N. Borozdin}

\affil{ NIS Division, Los Alamos National Laboratory, Los Alamos, NM 87545}
\email{gita@hea.iki.rssi.ru, kbor@lanl.gov, wpriedhorsky@lanl.gov}

\altaffiltext{1} {Also Space Research Institute, Moscow, Russia}

\begin{abstract}

Fast X-ray transients (FXTs) with timescales from seconds to hours 
have been seen by numerous space instruments. Because they occur at 
unpredictable locations, they are difficult to observe with 
narrow-field instruments. Only a few hundred have been detected, 
although their all-sky rate is in the tens of thousands per year.  
We have assembled archival data from Ariel-5, HEAO-1 (A-1 and A-2), 
WATCH, ROSAT, and \textit{Einstein} to produce a global fluence-frequency 
relationship for these events. 
Fitting the log N-log S distribution over several orders of magnitude
to simple power law we find a slope of $-1.0^{+0.2}_{-0.3}$.  
The sources of FXTs are undoubtedly heterogeneous,
representing several physical phenomena;  
the $\alpha\sim$ -1 power law is an approximate
result of the summation of these multiple sources.
Two major contributions come from gamma-ray bursts and
stellar flares.  These two types of progenitors are
distributed isotropically in the sky, however 
their individual luminosity distributions are both 
flatter than the -3/2 power law 
that applies to uniformly distributed standard candles.
Extrapolating from the BATSE catalog of GRBs, 
we find that the fraction of X-ray flashes 
that can be the X-ray counterparts of gamma-ray bursts 
is a function of fluence. The exact fraction of GRB-induced 
X-ray counterparts is sensitive to the $R_{X/\gamma}$ distribution, 
which we estimate from available experimental measurements.
Certainly most FXTs are not counterparts of standard gamma-ray bursts. 
The fraction of FXTs from non-GRB sources, such as magnetic 
stars, is greatest for the faintest FXTs. 
Our understanding of the FXT phenomenon remains limited and
would greatly benefit from a large, homogeneous data set, 
which requires a wide-field, sensitive instrument.

\end{abstract} 


\section{INTRODUCTION}

Essentially all missions sensitive to cosmic X-rays have detected intense 
X-ray outbursts with timescales from seconds to hours.
Peak fluxes range to 1 Crab and above 
(1~Crab=2$\times$10$^{-8}$ ergs~cm$^{-2}$~s$^{-1}$ in 2-10 keV band). 
X-ray outbursts are unlike classical X-ray transients, which persist 
for weeks or months. X-ray outbursts fade within a day and are normally only 
seen once.  Historically, X-ray outbursts seen for less than a day 
that lack persistent X-ray counterparts were called 
``fast X-ray transients''(FXTs).
Recently a new term, ``X-ray flashes''(XRFs), was coined for 
short (less than 1000 sec) intense bursts of X-ray emission\cite{heise01b}.
XRFs lack detectable gamma-ray emission, but are
reminiscent of gamma-ray bursts in their time history.
It can be difficult in some cases to distinguish
XRFs and other FXTs using archival data e.g. where the time resolution is inadequate. 

The sources of FXTs might be heterogenous; the experiments 
that measure them are certainly so. 
FXTs have been surveyed in a variety of energy bands, sometimes with 
low spectral and time resolution, and typically with low angular accuracy.
There 
is no straightforward way to obtain a uniform large set of FXT parameters. 
It is no surprise that different experiments have suggested different
sources for FXT, including flare stars, compact objects, extragalactic 
sources, and X-ray emission from GRBs [e.g. Pye \& 
McHardy 1983 \cite[hereafter][]{pm83}; Ambruster \& Wood 1986 
\cite[hereafter][]{aw86}; Castro-Tirado et al. 1999 \cite[hereafter][]{ct99}].
However, the quantitative contribution of these  
several sources is still unknown, and room remains for unknown phenomena. 
The discovery by BeppoSAX of long-lasting X-ray emission from GRBs, the 
so called X-ray afterglow \cite{piro98}, and the strong prompt  
X-ray emission from GRB seen by Ginga \cite{str98}
and WATCH \cite{sazon98}, shows that some fast X-ray transients are related 
to  GRBs. 


Theoretical work predicts the existence of GRB types that are 
X-ray rich or gamma-ray quiet. This might be caused by physical 
properties of the GRB central engine and
surrounding media, by the geometry of GRB source with 
respect to the observer's line of sight,
or by the redshift of classical (standard) GRBs (see recent review by \citealp{mesz02}).

In this paper we attempt to construct a uniform data set of FXTs
from the  heterogeneous data available to date. 
In section 2 we describe the experiments, and in section 3 our data analysis. 
In section 4 we discuss the connection of FXTs and GRBs, 
and estimate the fraction of GRB-related FXTs in FXT data. 
In section 5 we discuss other possible sources of FXT progenitors.

\section{EXPERIMENTAL DATA}

The importance of sensitive uniform sampling has been well recognized  
in the case of GRBs \cite[see][and elsewhere]{lee96,pet96}. 
The BATSE/CGRO experiment, with its large homogenous and uniform 
database, has allowed the study of statistical properties of classical GRB. 
A similar database, if it existed, would allow an unbiased determination
of the important parameters of FXTs, and 
could test different hypotheses for their origin. 
Although FXTs are by no means rare, we still do not have a large 
uniform set of these events.
In its absence, we have combined the data provided
by several very different experiments to quantify, as best we can,  
the statistics of FXT occurrence. While we recognize the imperfection of this
approach, it is the only way to estimate the FXT distribution over 
several orders of magnitude in fluence.
A brief description of the instruments and data sets is given below.

\subsection{Ariel-5}

\citealp{pm83} published a catalog of
FXTs detected in the 2-18 keV band by Ariel-5. 
Flux measurements from the rapid-spinning satellite were integrated in 
100 minute bins corresponding to the satellite orbit. These fluxes
provide a good  estimate of the integral fluence in 
the time bin. Transient events were selected by their detection 
in a single, or at most a few, satellite orbits. 
The  lightcurves therefore have 100-min time resolution.  
The limiting sensitivity was $\sim$20 mCrab (5.5$\sigma$) in 100 minutes. 
\citealp{pm83} found a cumulative log N-log F peak flux relation 
for 27 detected FXTs 
consistent with a power law with index $\alpha$ = -0.8 $\pm$ 0.5. 
\citealp{pm83} estimated the total number of FXTs above their
threshold as N=150-180 per year for the whole sky. 

\subsection{HEAO-1 A1 \& A2}

The HEAO-1 satellite carried two X-ray surveys, A1 and A2. 
These instruments scanned a great circle on the sky every 
35~minutes. Sources were sampled for a short interval once or twice
per scan
(10 sec for A1 and 60 sec for the A2 experiment).

\citealp{aw86} published 
a survey of transients detected by HEAO-1 A1 in the 0.5-20~keV band. 
This \citealp{aw86} survey was based on excess emission over an 12-hour interval. 
10 FXTs were detected, with a limiting sensitivity  about $\sim$4~mCrab. 
The log N-log F distribution was consistent with a power law with index 
$\alpha=-1.0\pm-0.5$.  \citealp{aw86} estimated the total number of FXTs as 
N=1500-3000 per year above their threshold, assuming that a 
typical event lasted more than 1.5 hours.

Connors et al. \cite[1986, hereafter][]{con86} published the  HEAO-1 A2 survey 
of FXTs in the 2-20 keV band. 
8 FXTs were detected with a limiting sensitivity  about $\sim$4-6~mCrab. 
The log N-log F distribution fit a power law with index 
$\alpha$=-1.0$\pm$0.7. The duration of the detected FXTs was poorly determined,
and might have  anywhere between 
60~s and $\sim$2000~s. The all-sky rate of such 
FXTs would have been between 10$^4$ per year for durations 2000~s and 
3$\times$10$^5$ per year for durations of 60~s. 

\subsection{WATCH}

\citealp{ct99} published the detection 
of 7 FXTs by the WATCH all-sky monitor onboard the Granat satellite. 
FXTs were detected in a 8-15~keV energy channel. 
WATCH, with a time resolution of several seconds, was able to
resolve the fast transients it detected ($\sim$100~min for 3 FXTs and 
$\sim$1~day for the rest). 
The typical peak flux for these sources was several hundred 
mCrab. Because the lightcurves are well-sampled, we have a good estimate 
of the total fluences. 
Sazonov et al. (1998) used WATCH to observe an additional 95 events in 
the energy band from 8 to 60 keV. 86 of them were observed by 
other GRB experiments, and confirmed to be GRBs, and 
47 events were localized to 1 degree or better. 
We used the Sazonov et al. (1998) data to constrain the
$R_{X/\gamma}$ ratio for gamma-ray bursts (see section 4).

\subsection{ROSAT}

Vikhlinin \cite[1998, hereafter][]{vikhl98} searched 
for faint X-ray bursts (duration 10-300~s) in pointed ROSAT PSPC 
observations with a total exposure of 1.6$\times$10$^7$ s. 
Only the 'hard' ROSAT energy channel (0.5-2~keV) was used for their study. 
A total of 141 X-ray flares with duration 100-300 s were detected.  
The quiescent flux was consistent with zero in approximately half of the burst 
sources, while positive quiescent emission was detected for the rest. 
112 bursts were identified as Galactic stars. One burst was from 
the X-ray binary LMC X-4. The 28 remaining FXTs were not identified, but for 
most of them starlike optical counterparts were found in 
the Digital Sky Survey.  The limiting fluence of these events 
(10-100~s duration) corresponds to 2.6$\times$10$^{-10}$~ergs~cm$^{-2}$ 
in the 0.5-2~keV band. We have translated the fluence from ROSAT energy 
band to 2-10~keV under assumption of Crab-like spectrum  
and  neglecting foreground absorption.  This implies a limiting 
fluence of 3.5~mCrab$\times$sec. The estimated rate of such events 
is $\sim$10$^6$ per year for the whole sky.   

Greiner et al. (2000) searched for GRB afterglows in the ROSAT all-sky survey.
During the survey, 
the telescope's field of view scanned a great circle on the sky,
sampling sources for 10 to 30~s.
The sensitivity limit  per scan is 10$^{-12}$ erg~s$^{-1}$~cm$^{-2}$.
Greiner et al. (2000) applied tight restrictions on the time profile
to extract GRB afterglow candidates, and selected 23 afterglow candidates.  
Closer examination  showed that about  
14 of afterglow candidates were in fact flares from late-type 
stars of class M-dMe.  With less restrictive 
requirements on the time profile 
Greiner et al. (2000) found more afterglow candidates,
mostly associated with flare stars.

\subsection{\textit{Einstein}}

Gotthelf, Hamilton \& Helfand \cite[1996; hereafter][]{ghh96} reported  
the detection of 42 faint X-ray flares in the Einstein IPC data. 
Those flares have typical fluxes from 10$^{-10}$~ergs~cm$^{-2}$ to 
10$^{-9}$~ergs~cm$^{-2}$ in the 0.2-3.5~keV energy range, and their durations 
were less than 10~s. Their detection rate corresponds to about 10$^6$ 
flares per year for whole sky. However, they concluded that only the hardest 
of their events, consisting of 18 flashes, was undoubtedly of
cosmic origin. The remaining soft events might have been detector artifacts.

\subsection{Ginga}

Strohmayer et al. (1998) reported on spectral fits and the 
$R_{X/\gamma}$ ratio for 22 GRBs observed with Ginga over 2 to 400~keV band.
We used their fits, together with data from other instruments, in 
our $R_{X/\gamma}$ distribution.

\subsection{BATSE}

BATSE GRBs are cataloged in the 4B and `Current' BATSE 
databases\footnote{http://gammaray.msfc.nasa.gov/batse/grb/catalog/}.
A recent catalog of bright BATSE GRBs was published by
Preece et al. \cite[2000, hereafter][]{preece00}.
Two searches by Kommers et al. (1999) and 
Stern et al. (2001) showed that the BATSE database contains 
almost as many untriggered GRBs as triggered GRBs. 
The untriggered GRBs tend to be fainter than the triggered GRB 
\cite[see Fig.8 of][]{stern00}, for natural reasons.  
We used both triggered and untriggered GRBs to obtain a log N-log S for GRBs
in the gamma-ray band. 

Spectral fits for 53 GRBs detected with BATSE were published by
Band et al. (1993).  Using the \citealp{preece00} catalog, we calculated
spectral parameters and estimated $R_{X/\gamma}$ for 81 more BATSE GRBs.

\subsection{BeppoSAX}

We used BeppoSAX afterglow observations \cite[see review in][]{front00a}
as inputs to for our estimate of the fraction of GRB-related FXTs. 
Frontera et al. (2000a, 2000b, 2001), in't Zand et al. (1999, 2000a, 2001), 
Pian et al. (2001) and Nicastro et al. (2001)
presented the X-ray and gamma-ray fluences for 15 GRBs in the 2-10~keV and 
40-700~keV bands measured by BeppoSAX.
We use these  data to help estimate $R_{X/\gamma}$.
Although we discuss the statistics of GRBs and XRFs reported
by Heise, in't Zand, \& Kuulkers (2000), and \cite{heise01,heise01b} 
we did not include them in our quantitative analysis because 
no numerical data for those XRFs have been published in useful form.

\subsection{RXTE ASM and Konus/Wind}

Smith et al. (2001) recently presented RXTE ASM and Konus/Wind
fluence measurements in the 1.5-12~keV and 50-200~keV energy bands
for 14 GRBs detected by both these instruments.
We have used these data for our $R_{X/\gamma}$ estimate.

\section{DATA ANALYSIS}

Our study started from the analyses published for the  
individual instruments.  We used both published results, and, in some
cases, data from public archives.  However, because
results of previous studies were strongly instrument dependent,
they cannot be directly compared. We need to convert data from the 
individual instruments into a quasi-uniform data set. 

Our analysis will proceed in two directions: in the forward direction
(approach \#1),
we will fit a power
law luminosity distribution (log N-log S) to the data.
In the backward direction (approach \#2) we
extract a model-independent luminosity distribution by plotting
each data set on the log N-log S plane. Approach \#2 requires  
the several assumptions discussed below.

Most of the aforementioned experiments measured the fluence 
(time-integrated flux) of detected events, but the HEAO-1 instruments 
provided only intermittent sampling of fluxes. These HEAO-1 results are 
important, however, because they were able to detect fainter FXTs than
Ariel-5 and WATCH.  ROSAT and Einstein detected even fainter FXTs,
but those experiments were sensitive in a much softer energy band, and 
are subject to additional corrections when converted to the 2 - 10~keV 
band. The WATCH data also need correction, because that instrument was
not sensitive to softer X-rays, but in this case the overlap
with other X-ray instruments is larger and hence
the correction is not so critical.  Additionally, interstellar 
absorption, which can be safely neglected in most cases above 2~keV, 
is an important factor in the ROSAT and Einstein bands,
adding more uncertainty to the results.  Finally, we did not have
information about individual FXTs detected with ROSAT and Einstein,
and used instead the integral estimates presented by \citealp{vikhl98} 
and \citealp{ghh96}.

For our primary analysis (approach \#1 above) we have used 
data from Ariel-5 (energy range 2 - 18 keV, 
minimum detectable fluence about 100 Crab$\times$sec), 
HEAO-1 A1 \& A2 experiments energy range 0.5 - 20 keV \& 2 - 20 keV, 
minimum detectable fluence 0.5 Crab$\times$sec), and WATCH (energy 
range 8 - 15 keV, minimum detectable fluence about 
several hundred Crab$\times$sec).
The fluence sensitivity limits above are given for long-duration events,
up to 1 day. The fluence sensitivity is much higher for short intense 
bursts, where background is not important.

To include HEAO-1 data into our analysis, we need to make 
assumptions about the duration and time profile of the transients.  
Fortunately, because of the nearly $\sim$1 slope of the cumulative 
distribution, almost the same distribution follows from a wide range 
of assumptions on duration. 
We have only lower (the scan duration) 
and upper (the scan interval) limits to the duration of the HEAO events. 
We estimated fluences for the HEAO-1 events by multiplying 
the measured flux by an assumed event duration. 
Our initial estimate of event duration was  taken as   
the geometric mean of the  minimum possible FXT duration (usually the 
scan duration of 10-60 sec) and the  maximum possible FXT duration, 
which is several hours for the A1 experiment and $\sim$4000~s
for the A2 experiment. 

To quantify the transient statistics, we assume that they
are distributed as
$N(>S)=A_0 \times S^{-\alpha}$, where $A_0$ is the 
normalization at 1 Crab$\times$sec, and determine the parameters of
this distribution with a maximum likelihood analysis. 
In our first analysis, we assumed
that the spectrum of the events was similar to the Crab nebula, 
i.e., a power law with photon index $\Gamma=-2$.
We then calculated the
minimum detection fluence for each experiment.
The overall event rates were taken from the published results of the original
analyses \cite{aw86,con86,pm83,ct99}. 

More detailed discussion of sky coverage, thresholds
and detection bands can be found in the references above. The parameters of our
maximum-likelihood determination of the parameters $\alpha$ and $A_0$ are 
shown in Fig.1.

It is known that many flare stars produce short, intense X-ray flares
\cite{stra93} with thermal spectra.
Because many FXT have been identified with
flare stars \cite{aw86,con86}, we have calculated the FXT distribution
that results if the FXT consist of mixture of Crab-like and thermal spectra
spectrum. A mixture of 30\% thermal-spectrum  FXT, with $kT=0.7 keV$, and 70\%
Crab-spectrum FXTs maximizes the maximum likelihood function, considering
the range kT = 0.5 - 4.5 keV, and `thermal' burst fraction 0 to 100\%.

Given this mix of FXT spectra, our best estimate of
$\alpha ; A_0$ yields $\alpha$=0.95$\pm$0.1 and $A_0=(4 \pm 1) \times 10^4$ 
at $1 Crab \times sec$. A conservative, 3$\sigma$ limit bounds the 
parameters to $\alpha$=[0.7-1.2], $A_0$=[$10^4$-$10^5$]
In any case, FXTs are common, occurring somewhere on the sky dozens to hundreds of 
times each day. They are one or two orders of magnitude more common than
classical gamma-ray bursts (Kommers et al. 1999).
Note that it is impossible to determine whether all the events that
make up this distribution fit our strict definition of occurring only
once. If some of the FXTs come from magnetic stars, RS CVns, or other
nearby objects, they may indeed repeat. These repetitions would 
have been seen by
HEAO-1, Ariel-5, etc. if those instruments had better sky coverage or longer
lifetimes.

Our distribution ( $\alpha=0.95$ and $A_0 = 4 
\times 10^4$, 70\% Crab-like vs.
30\% thermal spectra) is shown on  Fig.2. It is not significantly
different from the distribution that follows if we assume that 
all bursts have a Crab-like spectra. 

Fig.2 also shows the segments of the distribution seen by the individual
instruments (our analysis approach \#2). The shaded areas 
represent 1$\sigma$ errors for the individual experiment 
log N-log S curves, including the normalization and the statistics 
of the number of events seen by each experiment.

Although the duration of the HEAO-1 events is quite uncertain, this
uncertainty does not affect the overall distribution. We plot the
backward-derived log N-log S assuming the geometric-mean time duration
discussed above. But because the overall distribution has a slope
close to 1, changes in the assumed duration simply slide the 
log N-log S
contribution from the HEAO-1 instruments up or down along the same
curve. The dashed/dotted boundaries show the effect of changing the duration
to the extreme possibilities on the HEAO-1 A2 log N-log S distribution.

The ROSAT results from \citealp{vikhl98} and Einstein results from 
\citealp{ghh96} were not used to determine our net log N-log S distribution. 
However, they are plotted on Fig.2 for comparison.
We have translated the fluence measured in the ROSAT energy 
band to 2-10~keV by assuming a Crab-like spectrum with  
negligible foreground absorption.  The events reported 
by \citealp{vikhl98} had fluences from 3.5 to 200 mCrab$\times$sec. 
We estimated the total number of events 
per year for ROSAT events which had no detectable quiescent emission 
as $\sim$2$\times$10$^6$ year$^{-1}$, and for events  
without optical counterparts as $\sim$4$\times$10$^5$ events~year$^{-1}$.
We have taken these values as upper and lower limits to the FXT rate
at this fluence (shown as the cross on Fig.2)
The {\it Einstein} data were also converted to 2-10~keV assuming 
a Crab-like spectrum (shown as the rhomb on Fig.2), 
which is consistent with the spectral shape 
found by \citealp{ghh96} for hard events.

Our integrated distribution is consistent with a power-law 
with a slope $\alpha=-1.0^{+0.2}_{-0.3}$. 
A spatially homogenous distribution 
of identical sources (standard candles) would  have a slope -1.5. 
Our distribution is consistent with the slope of 
$\simeq$0.8 derived by PM83 from Ariel-5 events.
By adding data from several other experiments, we have extended
the PM83 distribution by several orders of 
magnitude towards the faint end.
The ROSAT and Einstein points, derived from \citealp{vikhl98} 
and \citealp{ghh96}, are consistent with an extrapolation of our
log N-log S curve.

Fig.3 shows the sky distribution of the events used in 
Fig.2 (Ariel, HEAO, and WATCH 
marked with crosses, and ROSAT and Einstein events marked with triangles).
The sky distribution is consistent with isotropy. 
The dipole moment of the distribution towards Galactic Center is  
0.07$\pm$0.08 , and the quadrupole moment is 0.31$\pm$0.04 (for an 
isotropic  distribution the dipole moment is zero, and the quadrupole moment 
is 1/3). We see no significant anisotropy for any subset of these data.
However we need to note that there are no reliable exposure maps for
mentioned above experiments, making it difficult to be definite about 
the lack of a disk population in these data.

\section{The Connection to GRB}

The connection between FXTs and GRBs is especially intriguing.
As a result of the Ginga and WATCH studies of prompt X-ray emission 
from GRBs \cite{str98,sazon98}, 
and the BeppoSAX discovery of GRB afterglows \cite{piro98}, it is evident 
that GRBs can emit a large fraction of their energy 
in the X-ray region \cite[see also][]{front00a}. 
This emission is both coincident in time with the gamma-ray
burst (prompt emission) and delayed (afterglow). 
Recently BeppoSAX has found a few GRB-like events with very little or no 
gamma-ray emission but strong X-ray emission
\cite{heise01,heise01b}.
These events were designated ``X-ray flashes''
\cite{heise01}.
However, Kippen et al. (2001, 2002) argue that they merge smoothly with
regular gamma-ray bursts in a single distribution
of their parameters. 
Such X-ray rich GRBs were estimated to total 
about 30\% of all GRBs \cite{kip02}.
Whether XRFs and GRBs 
originate from a single class of progenitors with a smooth
distribution of parameters, 
or are at least in some cases intrinsically different, is not yet clear. 
Comparison of GRBs and XRFs is fraught with selection effects, 
including the fact that the observed XRFs were selected 
in a different energy band (X-rays) than gamma-ray bursts.

In the absence of gamma-ray data, one might classify a 
conventional GRB that was detected in the X-ray band as an XRF. 
Grindlay (1999) examined Ariel-5 and HEAO-1 data
for an excess of gamma-ray quiet FXTs, above those expected
from conventional gamma-ray bursts, that might
indicate beaming of GRBs. He estimated that the GRB share of Ariel-5 and 
HEAO-1 FXT events was about 50\%.
We wish to refine this estimate by bringing several parameters
into our analysis.
These include the fraction of GRBs that emit in the X-rays, 
the ratio of the fluence emitted in the X-ray band to the fluence emitted 
in the gamma-ray band, the fraction of GRB that produce X-ray afterglows,
and the ratio of the X-ray fluence emitted in prompt phase to the fluence 
emitted in the early afterglow (the first $10^3 - 10^4$ sec)

BeppoSAX and WATCH \cite{front00a,sazon98} 
indicate that the first parameter is close to unity 
for long-duration GRBs. Long-duration GRBs are defined as events 
 with $T_{90} > 2$ sec in the BATSE data set
\cite{kouv93}.  The second 
parameter of interest is the ratio of the fluence emitted
in the X-ray band (2-10~keV) to the fluence emitted in the gamma-ray band 
(50-300~keV), $R_{X/\gamma}$. 
The ratio $R_{X/\gamma}$ can be estimated from the few events
observed in both the X-ray and gamma-ray band, or, using the larger
BATSE sample, by extrapolating the measured spectrum into the X-ray band.
If the latter is consistent with the former 
estimate, one could use the full BATSE GRB database to find 
the $R_{X/\gamma}$ distribution

Band et al. (1993) has shown that 
GRB spectra can be fit by the expression:
\[ N(E)=A\Big(\frac{E}{100keV}\Big)^{\alpha}exp\Big(-\frac{E}{E_{0}}\Big), \; \; (\alpha-\beta)E_{0}\ge E,\]
\[ N(E)=A\Big[\frac{(\alpha-\beta)E_{0}}{100keV}\Big]^{\alpha-\beta}\Big(\frac{E}{100keV}\Big)^{\beta}exp(\beta-\alpha), \; \; (\alpha-\beta)E_{0}\le E, \; \; (1)\]

in the 20 keV-2 MeV energy band. Using typical values for the parameters 
\cite[see for example][]{lloyd99,preece00}, 
the average value of $R_{X/\gamma}$ is about several percent. However, 
Preece et al. (1996) has shown that such an approach leads to an underestimate 
of X-ray fluxes for about 15\% of BATSE GRBs.  The magnitude of 
this underestimate can be relatively high (up to order of magnitude above 
Band's formula) for an individual GRB. 

Ginga, which was able to simultaneously measure X- and gamma-ray 
emission from GRBs, 
found an average ratio of prompt emission 
$R_{X/\gamma}$ of 0.24 for 22 GRB events, 
although the logarithmic mean $R_{X/\gamma}$
was 0.07 \cite{str98}. 
$R_{X/\gamma}$ was close to unity for several events. 
BeppoSAX confirmed that such a high ratio is not rare 
\cite{front00a, kip02}.

Data from different instruments are subject to significant 
differences in their energy bands and event selection criteria
and often do not agree between themselves.
To refine our estimate of $R_{X/\gamma}$
we have used all relevant available data
including BATSE, Ginga, RXTE, BeppoSAX and WATCH. 
Fits to the spectra of 53 BATSE and 22 Ginga GRBs, 
using the Band formula, are found in  Band et. al (1993) 
and Strohmayer et al. (1998).  
Using the catalog of BATSE GRBs by \citealp{preece00},
we have calculated the Band parameters for 81 additional GRBs
from BATSE data.
For all these bursts we calculated
$R_{X/\gamma}$ by integrating the Band formula. 
Smith et al. (2001) presented RXTE ASM and Konus measurements of 14 GRB
in 1.5-12~keV and 50-200~keV energy bands.
To convert ASM counts to X-ray fluence we fit the three-color ASM data 
to a power law for each GRB
(assuming a single power law in the 1.5-12~keV region). 
To obtain the 50-300~keV fluence from the single-channel Konus data
we assumed an average value for the Band parameters from \citealp{preece00}.
BeppoSAX data for 15 GRBs are presented at 
Frontera et al. (2000a,2000b,2001), in't Zand et al. (1999,2000a,2000b,2001), 
Pian et al. (2001) and Nicastro et al. (2001), 
who give X-ray and gamma-ray fluences for the 2-10~keV and 
40-700~keV bands. When available, we 
used the individual Band's parameters for each GRB, and otherwise used
average values from \citealp{preece00} for 50-300~keV channel.

Sazonov et al. (1998) published a catalog of 95 GRBs detected with 
WATCH.  We used 82 of these events, those for which fluences are available for
both energy channels (8-20~keV and 20-60~keV). Full data were unavailable
for the other 13 events. 
For the WATCH data we extrapolated $R_X$ and $R_{\gamma}$ 
to our 2-10~keV and 50-300~keV channels 
by assuming that the $\alpha$ Band's parameter was fixed at average value 
found from Ginga, RXTE and BeppoSAX measurements, $\sim$ -0.85, while the 
other  parameters were fixed at average values from \citealp{preece00}.

Fig.4 shows the distribution of $R_{X/\gamma}$ for all these events.  
The solid line shows the 
total $R_{X/\gamma}$ distribution derived from the Ginga, RXTE/Konus, 
BeppoSAX and WATCH measurements (134 GRB total), while the dashed line 
represents the BATSE events (134 GRB total). Note that the BATSE ratio
comes from an extrapolation of the measured spectra to the X-ray band.
The non-BATSE events, which are triggered at lower energies than BATSE, 
tend to have larger median $R_{X/\gamma}$, along with a tail of even higher
$R_{X/\gamma}$, X-ray-rich events. We will call them X-ray detected
GRB (XDGRB). The later distribution is in a good agreement with BeppoSAX
measurements \cite{heise01}.
We need to note that the $R_{X/\gamma}$ distribution certainly 
depends on how one weighs events  from the different experiments. 

We consider the XDGRB 
estimate of $R_{X/\gamma}$ to be more reliable than
the `BATSE' estimate, because the 
`BATSE' $R_{X/\gamma}$ ratio was calculated by extrapolation 
of Band's parameters from the 
20~keV-2~MeV region, while for the XDGRB 
$R_{X/\gamma}$ was actually 
measured, or calculated using Band's parameters 
fit to 2~keV-700~keV measurements.
The shift between the peaks of the two
distributions is in the direction expected from selection effects 
(see Appendix).  

To determine the integrated X-ray fluence from GRBs, we need to consider 
not only the prompt emission, but also the afterglow emission 
(the third and forth parameters mentioned at the beginning of this section). 
Data from 
BeppoSAX show that 80-100\% of GRBs that emit prompt X-ray emission also 
show an X-ray afterglow \cite{front00a}. We use this 
fraction (100 \%)as our estimate of the third parameter.

To estimate the fourth parameter (the ratio of the X-ray fluence emitted 
in the prompt phase to the fluence from the early afterglow) we 
use predictions of the standard fireball model.
The standard fireball model (Wijers, Rees, \& Meszaros 1997;
Sari, Piran, \&  Narayan 1998; \citealp{sari99})
predicts that in the afterglow phase, 
the X-ray flux decreases as a power law, as the GRB-induced blast waves 
sweep up surrounding matter and decelerate. The index of this power law
may change 
its value on time scales from minutes to hours. In this "external shock" 
model, both the tail of  the prompt X-ray emission and the afterglow 
are produced by the blast wave (Wijers, Rees, \& Meszaros 1997). 
The external shock model predicts that the amplitude 
of X-ray prompt emission during 
the final part of the GRB should be equal to the back-extrapolated flux 
from the afterglow. This back-extrapolation holds for 
most  GRBs that were observed by
BeppoSAX \cite{front00a}.

However, because the afterglow measurements by BeppoSAX start 6-8 hours 
after the GRB, there is no direct measurement of the X-ray 
flux immediately after the GRB. 
If the X-ray afterglow decays as a power law, its integral fluence can reach 
50\%-200\% of the prompt  emission \cite{front00a}.  
We have assembled from the literature decay indices for 15 GRB, and 
have calculated indices for 10 more, based on the published 
estimates of prompt and afterglow fluxes. 
The decay indices $\alpha$ fall in the range 
-1.0 to -1.9, most commonly about $\sim$1.3, and the distribution
of indices can be satisfactorily 
fit with a Gaussian.

In Fig.5, we combine all four parameters discussed above to obtain a 
log N-log S distribution for X-ray emission from GRBs. This 
estimate is based on the gamma-ray (50-300 keV) log N-log S 
distribution for all BATSE GRBs with duration $T_{90} > 2$~s. The solid 
curve is based on our XDGRB distribution of $R_{X/\gamma}$ ,  
sampled by Monte Carlo methods.  To the prompt X-ray 
fluences we add a  contribution from the afterglow obtained 
by Monte Carlo sampling of the afterglow decay parameter $\alpha$ and 
integrating the afterglow out to $10^4$ seconds. These curves plot 
our estimate of the X-ray fluence from cataloged, triggered BATSE events.

We also show for comparison the expected FXT distributions
for less-preferred distributions of  $R_{X/\gamma}$: the mean and logarithmic mean
of Ginga \cite[0.07 and 0.24,][]{str98} and our 'BATSE" distribution.
The less-preferred $R_{X/\gamma}$  yield
numbers of FXTs from GRBs that fall below the preferred estimate.

We would like to understand how these curves
would change if we included all the GRBs that are too faint to trigger BATSE. 
Some of them may be, in fact, X-ray rich GRB or XRFs (see \citealp{kip02}).
We can get some notion of this correction by adding to our sample the sample 
of untriggered BATSE events obtained by Kommers et al. (1998) and
Stern et al. (2000), which enhance the BATSE sample of the faint end. 
We find little change for all but the faintest part 
of the XRF log N-log S curve.
Fig.6 shows  the XRF log N-log S distribution for triggered, untriggered, and
all known BATSE GRBs, based on the XDGRB estimate of $R_{X/\gamma}$.

Fig.7 compares our prediction of the FXT's from GRBs to our
best-fit distribution of all FXT. The plotted distribution (solid curve)
is based on the XDGRB estimate of $R_{X/\gamma}$, and the population of all
(triggered plus untriggered) BATSE GRBs.

The distribution derived from the `BATSE'$R_{X/\gamma}$ (shown with dotted curve). 
BeppoSAX data (Heise, in't Zand, \& Kuulkers 2000, \citealp{heise01b})
suggest that about 70\% of FXT detected by BeppoSAX are 
GRB-related. The `BATSE' $R_{X/\gamma}$ would yeld a value of 0.1-1\% for this ratio, 
far less than observed.

Fig.7 also compares our estimate of the log N-log S curve
for FXT from GRBs with the rate of candidate GRB 
afterglows reported by Greiner et al. (2000). 
For any reasonable assumption about the $R_{X/\gamma}$ distribution, the 
Greiner et al. (2000) rate of events without optical counterparts
is in good agreement with our prediction,
suggesting that a substantial fraction of their residual, non-stellar, 
events are indeed GRBs.

It is clear that the fraction of fast X-ray transients
that can be attributed to 
classical gamma-ray bursts drops with decreasing fluence. 
The predicted  X-ray event rate for ``normal'' GRBs 
falls a factor of $\sim$~3-5 below the total FXT rate at high
fluences, but a factor of 100 below at the faint end. 
Even if we are off by a factor of two or three due 
to unknown effects, 
a large majority of FXTs are not counterparts of classical
GRBs (e.g, not X-ray flashes), especially at the faint end. 
However, we cannot rule out 
a possibility that exists a  population of X-ray rich GRB-like events 
that comprise a large fraction of the FXTs at the faint end.

\section{OTHER SOURCES of FXTs}

\subsection{Possible candidates}

The timescales and spectra of FXTs are diverse. Timescales range from seconds
to hours, and  spectra range from very hard \cite[kT$>$20 keV,][]{rap76} 
to soft \cite[kT$\sim$1~keV,][]{swank78}. \citealp{pm83} suggested the 
time profiles could be divided into classes, 
perhaps indicating multiple progenitor classes.
The few FXT identifications made to date confirm the 
heterogeneous nature of FXTs.

The lower right corner of our X-ray log N-log S curve is formed by several 
events with duration up to one day, which were 
seen mostly by WATCH \cite{ct99}. 
A T Tauri star was suggested by \citealp{ct99} as the progenitor 
for one of the events,
while the rest were probably generated by X-ray binaries.
Several X-ray binaries have been recently found to generate short,
powerful outbursts that might be classified as FXTs. For example, outbursts 
from V4641 Sgr up to  12.2~Crab (2-12~keV) with durations less than a day
were observed by the RXTE and BeppoSAX (Smith, Levine, \& Morgan 1999, 
\cite{zand00b,wij00})..
During the RXTE PCA observation, V4641 Sgr showed 
X-ray fluctuations by a factor of 4 on timescales of seconds and
$\backsim 500$ on timescales of minutes \cite{wij00}. 
The mass function identifies the system as a high-mass
X-ray binary harboring a black hole \cite{or00}.
A slightly longer type of X-ray transient, CI Cam, with a single short 
outburst about 2 days long, was also discovered by RXTE 
(\citealp{smith98}, Revnivtsev, Emel'yanov \& Borozdin 1999).

``Super type I X-ray bursts'' that last from 30 min to 3 hours 
 were reported recently from several low-mass
X-ray binaries \cite[e.g.,][]{cor00,wij01}.
Similar events might been detected as high-fluence FXTs by low-resolution 
experiments.
These or similar sources might contribute to the lower right corner 
of the FXT log N-log S curve. 

In't~Zand et al. (1998), and Kaptein et al. (2000) announced the 
discovery of type I X-ray  bursts from `empty' places. These bursts, 
observed from locations without steady emission, would be 
naturally classified as FXTs. The origin of such
bursts might be low accretion rate neutron star binaries. 
We can get an idea of the rate
of `empty' place bursts from BeppoSAX\cite{zand01_1}.
It observed the Galactic Center region for $T_{obs}=4\times 10^6$~s 
and detected about 
$\sim$1500 X-ray type I bursts from  $N_b$=31 sources. 
At the same time it observed $B_e$=4 X-ray type I bursts from 
4 `empty place' X-ray bursters. If all these events were generated
on low-accretion X-ray neutron stars (with, say, X-ray luminosity less 
than several percent of the Eddington limit), and their space distribution 
follows the space distribution of normal X-ray bursters, we can
estimate the total number of bursts from `empty place' per year per sky
as: $N_{sky}= (1_{year}/T_{obs}) \cdot (N_t/N_b) \cdot B_e \sim 50$ events,
where $N_t\sim$~50 is the total number of known normal bursters. If their 
distribution is isotropic, the number of bursts from `empty' places would be 
at least an order of magnitude greater.This simplified estimate 
agrees well with the 
BeppoSAX estimate \cite{cor02a}. They found that the total number of such 
sources in our Galaxy is between 30 and 4000, while the recurrence time between 
the bursts is a half year (for 30 sources) and several 
tens of years (for 4000 sources)
The fluence from such events would be  comparable with outbursts from 
RS CVn binaries, yielding
about 50-100 events per sky in the $10-1000 Crab \times sec$ region.

\citealp{pm83} identified a large fraction (6 out of 27) 
of Ariel-5's fast X-ray transients with RS CVn binary systems. 
Another RS CVn object was proposed as a possible FXT source by \citealp{aw86}. 
RS CVn systems are binaries formed by a cool giant or subgiant with 
an active corona and a less massive companion in a close synchronous  
orbit. RS CVn binaries typically exhibit a peak flux 
$\sim$10$^{32}$ erg/s for a duration of 1 to 10~h. 
Outbursts from nearby RS CVn systems would have a fluence that puts them in 
the middle part of our diagram. This is certainly true for the identifications
proposed by \citealp{pm83} and \citealp{aw86}.

Our analysis of GRB X-rays indicates that this contribution to the FXT 
log N-log S is most significant around 100-1000 Crab$\times sec$.
Heise et al. (2000) reported that about 70\% of all FXTs detected with
BeppoSAX are GRB-related,
i.e., either conventional GRBs or XRFs.  1
Our best estimate (XDGRB) indicates that the GRB 
contribution to FXTs (see $R_{X/\gamma}$ distribution discussion) 
peaks at 20-30\%.  However, taking into account
the uncertainties of both studies, we consider these results as
moderately consistent between themselves, and consistent 
with \citealp{grin99}. 
We note that BeppoSAX/WFC detected $\sim$1.5$\times$10$^3$ thermonuclear 
X-ray bursts from $\sim$35 sources, but did not classify them as
FXTs because they had clear identifications. 
Some of these bursts might have been identified as FXTs by earlier experiments
like Ariel-5. This would increase the total number of FXTs and decrease
the GRB fraction.  Our
estimate of GRB-related FXTs is also consistent with the ROSAT 
result \cite{grein00} at low fluences.  We have shown 
(see Fig.7) that the contribution of classical BATSE GRBs drops 
dramatically for lower fluences. Other sources must dominate
FXT statistics that region.

\citealp{con86} found 6 of their 10 FXTs to be flares from late 
type dMe-dKe stars, i.e., M or K dwarfs with Balmer lines in emission. 
\citealp{aw86} also identified 3 of 10 HEAO-1 A1 FXTs with flare stars.
Flares of dKe-dMe have peak fluxes of 10$^{28}$-10$^{32}$~erg~s$^{-1}$  
and duration from $\sim$minutes to $\sim$hours.
Their spectra have 
temperature $\sim$1~keV,
and they should be well detected by ROSAT.  Indeed, 
\citealp{vikhl98} found star-like 
counterparts for 132 out of 141 reported FXTs. We therefore suggest 
that the least luminous and most numerous 
part of log N-log S distribution is formed by nearby flare stars.

There is however a fraction of lower-fluence FXTs that are
not identified with star-like objects.  While the contribution
of classical GRBs should be negligible in this region, 
other types of progenitors may play a more important role.
One of them, gamma-ray quiet GRBs might arise in several 
ways \cite[e.g. ][]{macf99,mesz00}.
Recent BeppoSAX results have shown that
some GRBs are faint in the standard 
50-300~keV BATSE band but bright in X-rays
\cite{kip01,kip02}.
Out of the 53 GRB-like transients detected with the BeppoSAX WFC, 
17 were not registered by BeppoSAX GRB monitor, 
which is sensitive in 40-400~keV energy region. \citealp{kip01}
found that 9 of such XRFs were also seen 
by BATSE and recorded as untriggered events at BATSE database.  
According to Sazonov et al. (1998) 10\% of WATCH events 
had detectable 
emission in 8-20 keV energy channel but no significant emission in 
the hard channel.  Our analysis (see Fig.6) confirms 
that untriggered BATSE events contribute only modestly to the lower fluence 
part of log N-log S if they follow the same $R_{X/\gamma}$ distribution
as the triggered events.  But GRB-like outbursts with softer spectra
might be more significant at lower fluences.

Other exotic progenitors may contribute also. 
Schafer, King, \& Deliyannis 2000, suggested that ordinary main sequence 
stars like our Sun may exhibit rare powerful superflares with
$10^{33}$ - $10^{38}$ ergs, some of them in X-rays. 
However, one cannot quantify their contribution until one better understand the
frequency of such superflares.

Extragalactic sources may also generate FXT phenomena. X-ray variability 
is a fundamental property of Active Galactic Nuclei. In the last decade it was 
found that BL Lac objects (blazars) exhibit strong correlated variability 
in both X-ray and TeV gamma-ray bands on short 
timescales of days to hours \cite{cat97,marachi99}.   
The brightest of such events may be detected as FXTs.

\subsection{Contributions to the Fluence Distribution}

Previous searches for FXT progenitors
demonstrated that we probably deal with a mix of close (flare stars) and
extragalactic (GRB-related) events.
The sky distribution for both these populations is expected to 
be isotropic. This is consistent with the sky distribution of detected FXT 
(Fig.3).
However, the absense of reliable exposure maps for
many observations and possible selection effects leave room
for the existence among FXTs of
a non-isotropic population with the Galactic disk or bulge distribution.
The slope of the log N-log S distribution of such additional component 
is expected to be flatter than -1.5 for a variety of possible progenitors.

The most natural candidates for an additional component of FXTs
are low mass X-ray binaries (LMXBs). 
\citealp{cor02b} have analyzed BeppoSAX data and found several 
X-ray type-I bursts from 'empty' places, where no persistent 
emission was detected down to a several mCrab limit. 
\citealp{coc01} have proposed a new class of low-luminosity
bursters. \citealp{cor02b} have shown that all detected bursts 
from 'empty' places are indeed the members of such 
class and follow LMXB space distribution.  
Since LMXBs are concentrated in the Galactic Bulge 
(see e.g. recent review by Grimm, Gilfanov, \& Sunyaev 2002),
any associated extra FXTs should be clustered 
around the Galactic Center.
However, most early surveys had difficulty 
in detecting FXTs from this part of the sky \cite{war81}. 
Their localization accuracy was poor (typically several degrees), 
and any X-ray flash from 
a populated region of the sky would tend to be 
attributed to known X-ray sources. 
Hence, the population of LMXB with low persistent flux 
and emitting rare bursts would 
need to be 
be quite numerous to have been clearly detected by those instruments. 
Emel'yanov et al. (2001) have shown that TTM observations of the
Galactic Bulge put strong constraints on the number-frequency function of 
low-accreting X-ray bursters indicating that
the number of low-luminosity X-ray bursters $\times$ the burst 
frequency from a burster is small. 
If the FXTs that are distributed as a bulge population are fewer than several
percent of all FXTs, they cannot be found with the usual statistical tests
(e.g., a K-S test would not identify a bulge fraction less than 3\% of
the total as significant).
The deviation from uniformity in the FXT distribution would be
further masked by the poor ability of previous studies to find FXTs at low
latitudes \cite[see e.g.][for the discussion of source confusion in Ariel-5 
survey]{war81}.

Our two most reliably identified
source populations, gamma-ray bursts and 
stellar flares, although isotropically distributed, might yield a
log N-log S distribution with a slope significantly flatter than 
-3/2 (as would be expected for a uniform distribution of 
standard candles). In the
case of gamma-ray bursts, the distribution is known to flatten due
to cosmological effects. 
Even more important than the cosmological flattening, the luminosity
distribution of FXTs from GRBs is flattened by the broad distribution of $R_X/R_{\gamma}$. 
We demonstrate below that nearby flare stars 
would also produce a distribution flatter than -3/2.

EXOSAT observed 22 late-type flare stars 
(dMe-dKe) in the vicinity of the Sun (2-20 pc) 
in the 1-10~keV energy region. 
They found that these stars generated X-ray flares with total energies
from 10$^{30}$ to 10$^{34}$ ergs and $N(>E) \sim E^{-0.7}$ 
(Pallavicini, Tagliaferri, \& Stella 1990), e.g. with a
differential frequency-flare distribution
fits a power law N(E)$\sim E^{-1.7}$.  
A similar 
distribution function has been observed for the Sun, with  
differential frequency-flare energy distribution index
$\Gamma=1.8$ \cite{ger89} or 2.0 \cite{ver02}, and for optical flares 
on the M dwarf flare star AD Leo (Pettersen, Coleman \& Evans 1984).

The integral distribution of X-ray flares from flare stars is:

\[ N(>S) \propto \int\limits_{r = R_{min}}^{\infty}\int\limits_{E = 4 \pi r^2 S}^{\infty} 
n(r,E) dE dr, \; \; (2)\]

where $n(r,E) \propto r^2 \times E^{-\gamma}$ is the differential 
distribution function of X-ray flares from flare stars, 
E is the flare energy, 
$R_{min}$ is the nearest flare star,
and S=E/(4 $\pi r^2$) is the measured fluence. 
This distribution holds for sources uniformly distributed through the space 
(as is certainly within a radius of 100 pc), with a differential frequency- 
flare energy function described with power law  $E \propto E^{-\gamma}$. 
We derive:

\[ \int\limits_{E =  4 \pi r^2 S}^{\infty} E^{-\gamma} dE =  
 (4 \pi r^2 S)^{-\gamma+1} \propto r^{2-2\gamma} \times S^{1-\gamma}   \; \; (3)\]

substituting $\gamma = \alpha +1$, where $\alpha$ is the slope of the 
cumulative frequency-flare energy function, we get:
 
\[ N(>S) \propto S^{-\alpha} \times \int\limits_{r = R_{min}}^{\infty} r^{2-2 \alpha} dr \; \; (4)\]

If the total flare energy  dynamical range 
$E=E^{-\gamma} [E_{min},E_{max}]$, is greater than the distance 
dynamical range $R^2 [R=(R_{min},R_{max})]$, 
then for a range of detected fluences 
$S=[E_{min}/R_{min}^2,E_{max}/R_{max}^2]$ the slope of 
the detected cumulative number-fluence function 
$(N(>S))$ will be the same as the slope of the integral 
frequency- flare energy function $N(>E) \propto E^{-\alpha}$ 
  
EXOSAT data confirm that $N(>E) \propto E^{-0.7}$  for at least  
$E_{max}/E_{min}=10^4$ (Pallavicini, Tagliaferri, \& Stella 1990).  
So, for example, if $R_1^2/R_2^2=100$, the dynamical range where 
$N(>S) \propto S^{-0.7}$ is $S_1/S_2=100$.

Integrating over the full range of distances (2-100 pc) and
flare energies ($10^{31}$-$10^{34}$), we estimate that the integral 
fluence distribution 
$(N(>S))$ from stellar flares has a slope that steepens slowly from 
$-\alpha$ to -3/2 with increasing fluence (Fig.8).
Note that over 3 orders of magnitude in fluence, it 
is close to a power-law with slope $\sim -0.6 \approx -\alpha$.

The intrinsic distribution of flare energy drives 
the observed log N-log S fluence 
distribution for the stellar component of FXTs. The fluence 
(Crab$\times$s) 
in the 2-10~keV region is related to the flare energy as $3\times 10^{30}$ 
(erg) $\times$ M (at Crab$\times$sec) $\times$ $D^2$ (at $pc^2$).  
The HEAO-1 experiments had a fluence detection 
threshold about 1 Crab$\times$s, while Ariel-5 had a threshold of 
about 100 Crab$\times$s.  As a result, HEAO-1 detected 
flares with energy $L\geq 1 \times 10^{31}$ ergs
from sources located within 2 pc, and
the most energetic flares from distances up to 100 pc. Ariel-5
was able to detect flares with $L\geq 1 \times 10^{33}$ 
within 2 pc, and assuming the most energetic flares within 10 pc. 
Taking the star density near the Sun $\sim$0.1 star $pc^{-3}$, 
and assuming,  based on the standard 
4-component model of the Galaxy \cite{bahc86},that stars 
are distributed uniformly within 100 pc of the Sun, HEAO-1 
should have seen  flares from
up to $4 \cdot 10^5 \times K_{fs}$ flare stars, 
while Ariel-5  was able to detect $400 \times K_{fs}$ flare stars.  
The catalog of chromospherically active stars 
\cite{stra93} includes 3 stars closer than 10 pc, while
Pallavicini, Tagliaferri \& Stella (1990)
observed 12 flare stars within 10 pc.
Basing on these data we accept $K_{fs}=2.5 \cdot 10^{-3}$  
as a lower limit for fraction of dMe-dKe stars around the Sun.  
In this case the HEAO-1 experiments were able to detect 
X-ray flares from $10^4$ flare stars. 
Because the HEAO-1 experiments were sensitive to fainter events than
Ariel-5, and because flare star FXTs are generally fainter than
RS CVn FXTs, we expect that the ratio between late-type flare stars
and RS CVns should increase as we move from the Ariel-5 to the HEAO-1
fluence band. 
This is consistent with their proposed identifications:
HEAO-1 identified 10 events with dMe-dKe stars, while Ariel-5  
reported only one such identification. 

In summary, over a wide range of fluences, the intrinsic distribution 
of flare energies dominates over the
geometric term, so that the -3/2 power law expected for standard
candles does not apply. The shallow slope of the FXT distribution is,
therefore, obtained naturally even for uniformly distributed
progenitors.

\section{SUMMARY}

We have produced a composite log N-log S relationship for 
fast X-ray transients, by combining data
from several experiments. This relation spans 4 orders 
of magnitude, and is fit  to a power law
with a slope of $-1.0^{+0.2}_{-0.3}$. 
Extrapolation of this slope to lower fluences 
is consistent with the sensitive ROSAT and Einstein experiments.
The sources of fast X-ray transients are undoubtedly heterogeneous.  
The $\alpha\sim$ -1 power law must
come from  the summation
of contributions from several different progenitor classes.
Both GRBs and flares stars will contribute components to this distribution 
that are flatter than the `isotropic' slope of $\alpha = -3/2$.
The exact shape of the global fluence-frequency relationship 
might deviate from this simple power law, even inside the range we measured. 
However, due to the small number of
events that we analyzed, and resulting large uncertainty in the deduced rate,
a more complicated luminosity function is not justified
(see for example detailed discussion at Murdoch, Crawford, \& Jauncey (1973)). 

We use archival measurements of the X-ray and gamma-ray 
fluences for 134 GRB to estimate the distribution of the X-ray 
to gamma-ray ratio. This $R_{X/\gamma}$ distribution shows the
presence of highly significant tail of X-ray rich events.
Extrapolating from the BATSE GRB catalog, 
we find that the fraction of fast x-ray transients 
attributable to classical gamma-ray bursts 
is a function of fluence, being greatest for high fluences and dropping
dramatically at the faint end.
The exact fraction of GRB-associated fast X-ray transients
is sensitive to our assumptions about
the $R_{X/\gamma}$ distribution, but 
 our analysis certainly shows that most of fainter
FXTs are not counterparts of classical gamma-ray bursts.
The fraction of FXTs from non-GRB sources, such as magnetic flares
on nearby stars, is the highest for the faintest flashes. 

Our understanding of the fast X-ray transient phenomenon is still
poor. The situation is similar to that for gamma-ray bursts 
before BATSE.  There have been many FXTs detected, but by a variety
of instruments each with specific restrictions.
The data show  convincingly that FXTs  are ubiquitous.
But the number of events detected by each
particular experiment is small, and their statistical analysis is
complicated.  We can identify some classes of progenitor sources, but
we can determine the fractional contribution of these classes only roughly.
There remains room for the discovery of new types of sources that 
produce fast X-ray transients, but
we certainly cannot yet confirm the existence of new classes like 
``gamma-ray quiet GRBs''
or ``superflares'' at normal stars. A large uniform dataset
from a dedicated, sensitive wide-field X-ray experiment 
\cite[see e.g.][]{lobster,pried00,bor99}
could revolutionize our understanding of the FXT phenomenon. 
The greatest opportunity for discovery may lie in the faintest 
FXTs, posing a challenge to instrument builders.

\acknowledgments

The GRB data were taken from the BATSE archive 
{\footnote {\url http://gammaray.msfc.nasa.gov/batse/grb/catalog/}}.
The work of VA was partially supported by  Russian Foundation for 
Basic Research grant 01-02-17295.
KB is glad to acknowledge helpful discussions of
the subject with Jean in't Zand and John Heise.
VA would like to aknowledge the advices from Eugene Churazov and 
Marat Gilfanov.

\appendix

\section {Appendix. Selection effects on $R_{X/\gamma}$ ratio}

Differences in the sensitivity band and observational
strategy significantly affect the
$R_{X/\gamma}$ ratio measured with different instruments.
In our analysis we made use of all data we aware of to obtain the most
instrument-independent estimate possible.
We calculated the  mean, median and variance of the measured distributions
of $R_{X/\gamma}$ for the `BATSE' and XDGRB samples (Table 1) 
We also fitted the distribution $R_{X/\gamma}$ for the XDGRB and `BATSE' 
samples with a Gaussian, and  with a Gaussian plus quadratic polynomial. 
These functions were choosen simply to match the data with a smooth curve. 
A Gaussian alone cannot fit the tail of distribution. The difference
between the Gaussian cores of the distribution is in the direction
expected from the triggering effects, but the tail in the XDGRB distribution
cannot be caused by triggering effects.
The inset on Fig.4 shows the best-fit curves for the Gaussian plus
quadratic polynomial fit to the both samples, to show qualitatively how they 
differ.

To understand the effect of triggering on the $R_{X/\gamma}$ distribution,
we consider the simplified case in which the X-ray/gamma-ray fluence ratio 
follows a log-normal distribution. In other words, if $\alpha$ is defined 
as $F_x \tbond exp(\alpha) F_\gamma$,  $\alpha$ is distributed as a Gaussian.
The Gaussian distribution of $\alpha$ is 
dN/d$\alpha\propto exp[-(\alpha-\mu)^{2}/2\sigma^{2}]$, where $\mu$ is 
the mean of the distribution and $\sigma$ is its standard deviation, for 
events triggered in the band of $F_\gamma$.

If the log N-log S distribution goes as $N\sim S^{-\beta}$, 
then the X-ray triggered distribution of  $\alpha$, for the same population 
of events, will follow a Gaussian distribution of the same width, 
but centered on $\mu(X_{ray}) = \mu(\gamma) + \beta \cdot \sigma$. 
We assume that the X-ray trigger band is the same band as for $F_X$.
In the case of an isotropic, $\beta = 3/2$ power law, the distribution 
will be displaced 1.5 standard deviations to the right when we trigger 
on X-rays instead of gamma rays.  
For trigger bands other than those used for $R_X$ or $R_{\gamma}$, the 
displacement will be different. 

This selection effect may explain the fact that WATCH 
generally selects softer GRBs than BATSE. While BATSE triggers 
in the 50-300 keV band, WATCH triggers in the 8-100 keV band. 
However, the nominal trigger range for Ginga, at 50-400~keV is harder 
even than BATSE.  The softer $R_{X/\gamma}$ ratio 
obtained by Ginga may be the result of a secondary selection 
done by Strohmayer et al. (1998). While Ginga observed about 
120 GRBs \cite{ogas91,fenimor93}, only the GRBs with good quality 
spectral data in both $\gamma$- and X-ray detectors were selected 
for spectral analysis, possibly driving the selection of softer,
X-ray bright, GRBs. 

We can make a rough check of the selection effect by taking the power law 
$\beta$ from BATSE measurements, calculating the displacement 
$\beta \cdot \sigma$, and
comparing it with the displacement between the BATSE and the other,
mostly X-ray triggered, $R_{X/\gamma}$ distributions.
We predict a  displacement for $\beta \sim$~1.3 (derived from middle part 
of BATSE GRBs log N-log S) of 0.6-0.7, compared to a displacement in range of 
0.5-0.7 (Table 1), between the $R_{X/\gamma}$ distributions for
`BATSE' and  XDGRB. 
The direction is as expected: triggering at softer energies 
moves the $R_{X/\gamma}$ distribution to larger values.

We have tested whether the difference between the `BATSE' and XDGRB 
distributions was caused by the effect of trigger-band selection 
on a single distribution of GRBs, 
or if there existed an additional X-ray rich GRB population 
not detected by BATSE. To do this we have fit the $R_{X/\gamma}$ 
distribution of the BATSE GRB sample to a log-normal 
distribution. A maximum-likelihood estimate of the fit parameters 
gives $\mu(X_{ray}) \approx 0$ and $\sigma = 0.5$. 
This result shows that except of the triggering band effect
there is also an additional contribution of X-ray rich events
for XDGRB case 
{\footnote {~See also discussion at \citealp{kip02}}}.

\newpage

\begin{figure}
\plotone{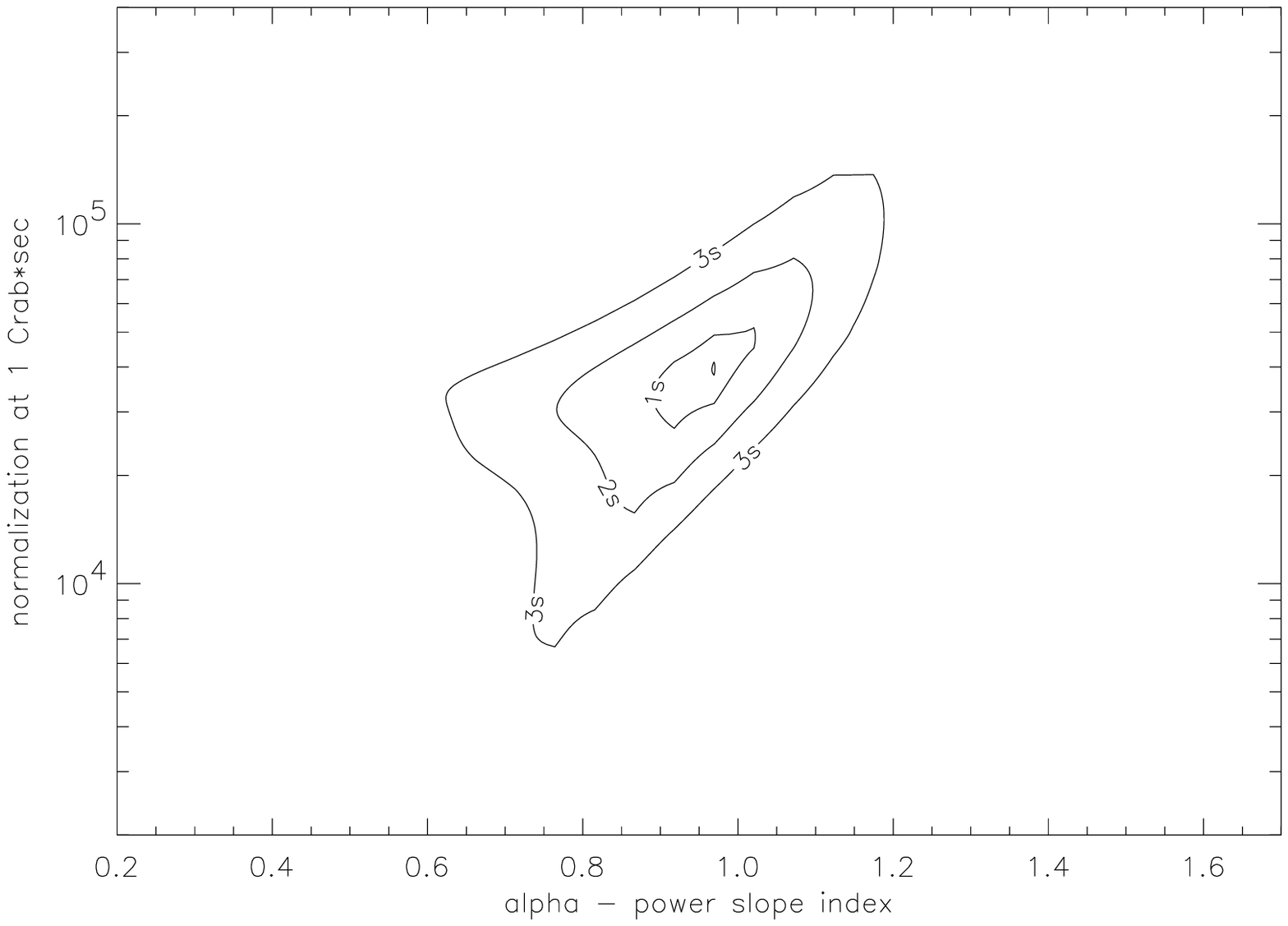}
\caption{Maximum-likelihood confidence limits for the parameters of
a $N(>S)=A_0 \times S^{-\alpha}$ power-law distribution of FXTs.
The contour lines show 1, 2, and 3 $\sigma$ significance levels.
}
\label{parmaxl}
\end{figure}

\begin{figure}
\plotone{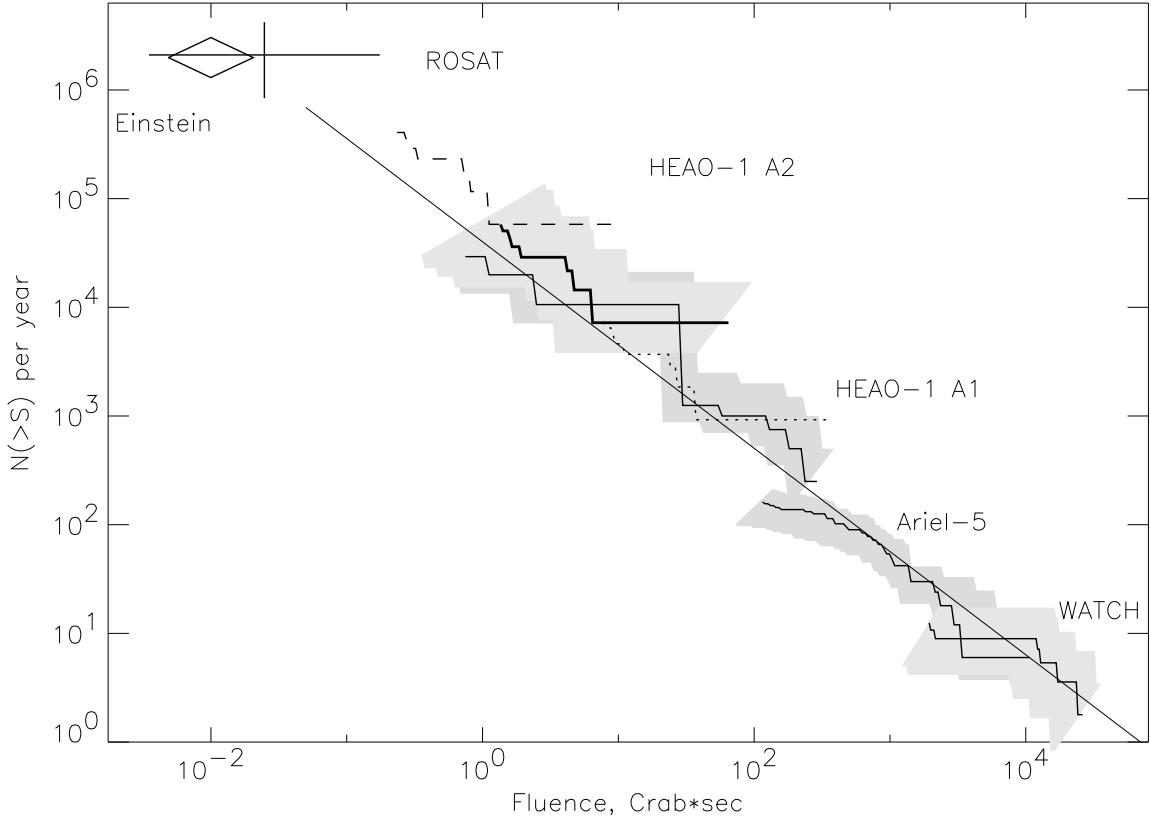}
\caption{Observed log N-log S distribution for FXTs, assuming
a power law fluence distribution 
($N(>S)=A_0 \times S^{-\alpha}$).
The thick solid line is best-fit power law derived from a maximum
likelihood fit to 
Ariel-5, HEAO-1, and WATCH data, with
 $\alpha=0.95$ and $A_0 = 4 
\times 10^4$ 
The shadowed areas show the measurements of the log N-log S distribution
from each experiment, including both event statistics and systematic
errors. 
The cross denotes  ROSAT events \cite{vikhl98}, 
and the  diamond  Einstein FXTs 
The dashed and dotted curves show the range of distribution that follow from 
the HEO-1 A2 events, given extreme possibilities for event duration.
\cite{ghh96}
}
\label{logns}
\end{figure}

\begin{figure}
\plotone{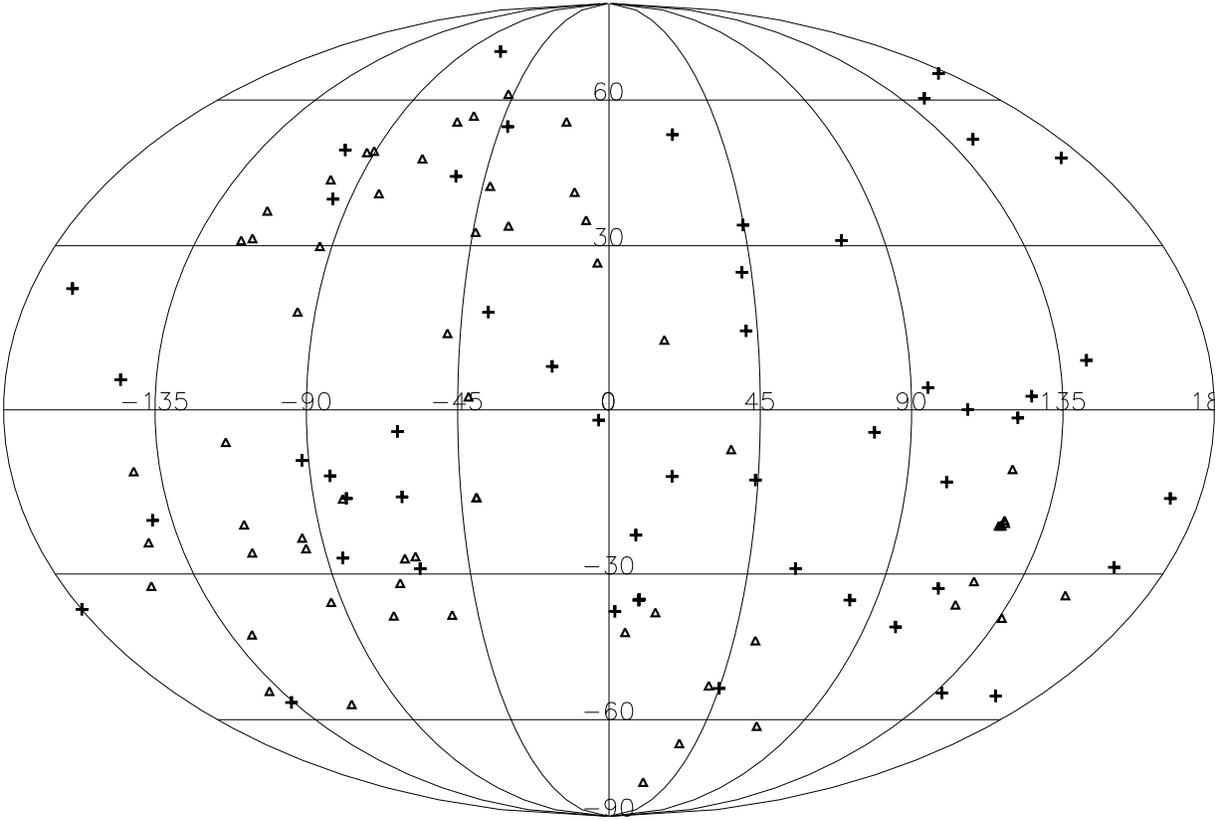}
\caption{Sky distribution of the FXTs included in the log N-log S distribution 
of Fig.2.
The crosses are  the Ariel-5, HEAO-1, and WATCH events; triangles are
the ROSAT and \textit{Einstein} events
}
\label{sky}
\end{figure}

\begin{figure}
\plotone{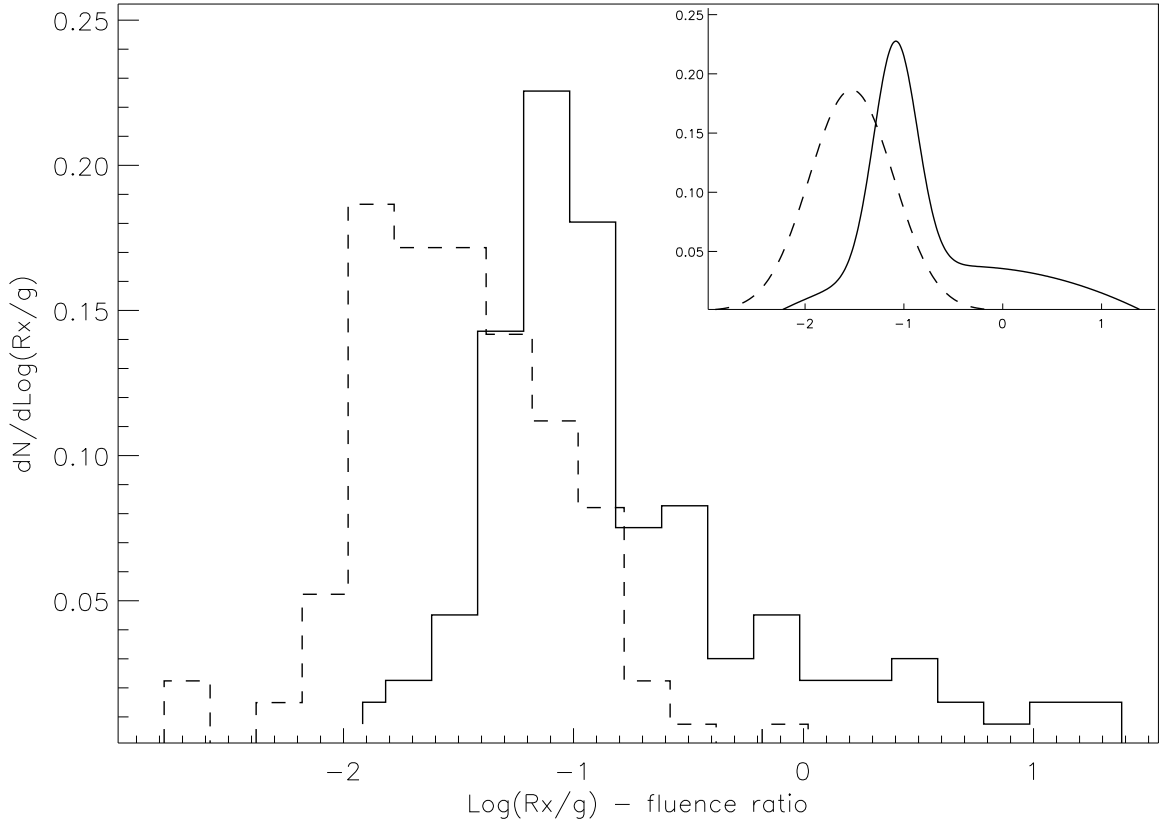}
\caption{
$R_{X/\gamma}$ distribution for GRBs. (2-10~keV fluences /50-300~keV 
fluences). The solid line is from Ginga, RXTE/Konus, BeppoSAX, 
and WATCH events (XDGRB case), and the dashed line from BATSE events. 
The former were measured in the X-rays, but the latter were extrapolated 
from the BATSE band (above 20~keV) to the X-rays.
The inset shows the best fit to these distributions, using a Gaussian 
plus a quadratic polynomial (fit parameters are given at Table 1). 
The solid curve is the XDGRB sample, and the dashed curve is the  
`BATSE' sample.
}
\label{rxrg}
\end{figure}

\begin{figure}
\plotone{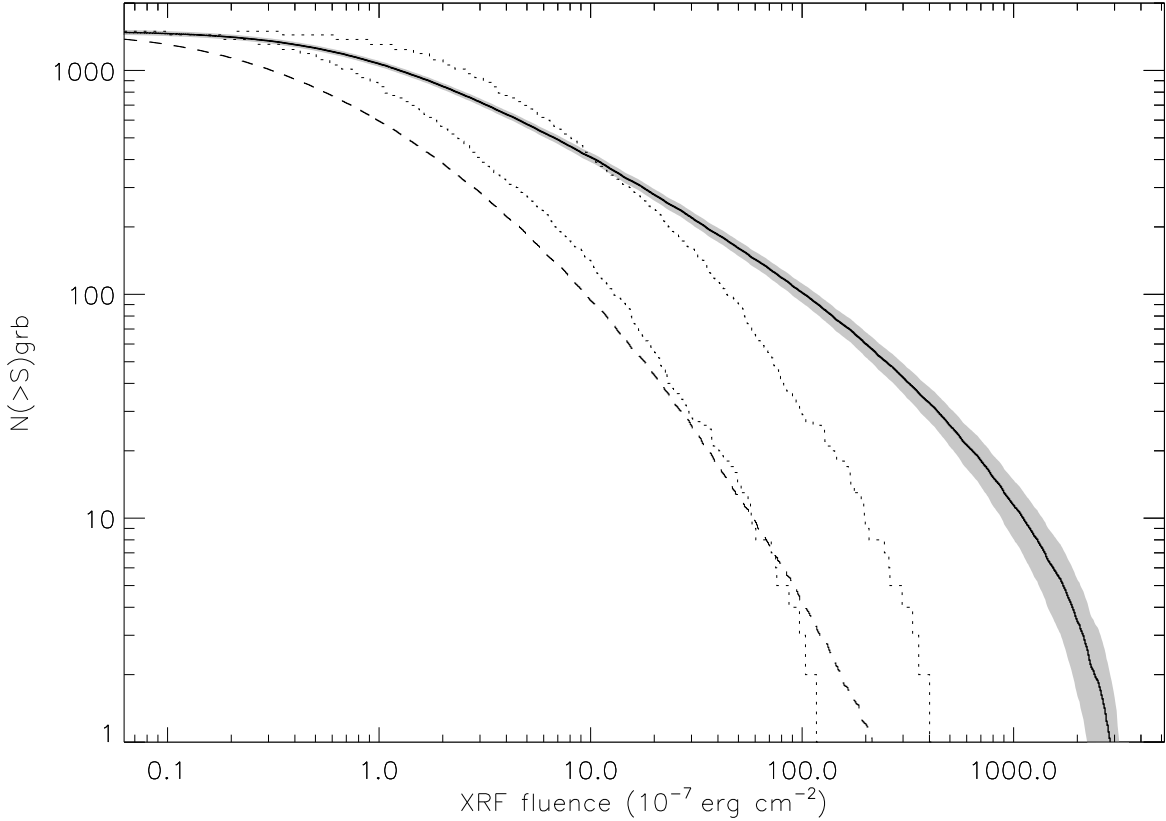}
\caption{
The expected log N-log S X-ray distribution (summing
prompt and afterglow emission) for BATSE triggered GRBs,
based on different distributions of the
$R_{X/\gamma}$ ratio.
The thick solid curve is based on the XDGRB distribution $R_{X/\gamma}$, derived 
from Ginga, BeppoSAX, RXTE/Konus and WATCH events (solid line in Fig.4). 
The shadowed
area shows $1\sigma$ errors for this solid curve, based on Monte Carlo
sampling of the $R_{X/\gamma}$  distribution.
The dashed curve is based on  a $R_{X/\gamma}$ derived
from BATSE spectra (`BATSE' case).
The error region for the `BATSE' case has a similar size as for 
the XDGRB case.
Finally, the dotted curves shows the expected  log N-log S distribution for the 
fixed values of 0.07 and 0.24 found by Ginga
(their logarithmic and arithmetic averages; \citealp{str98}).
}
\label{grb_logns}
\end{figure}

\begin{figure}
\plotone{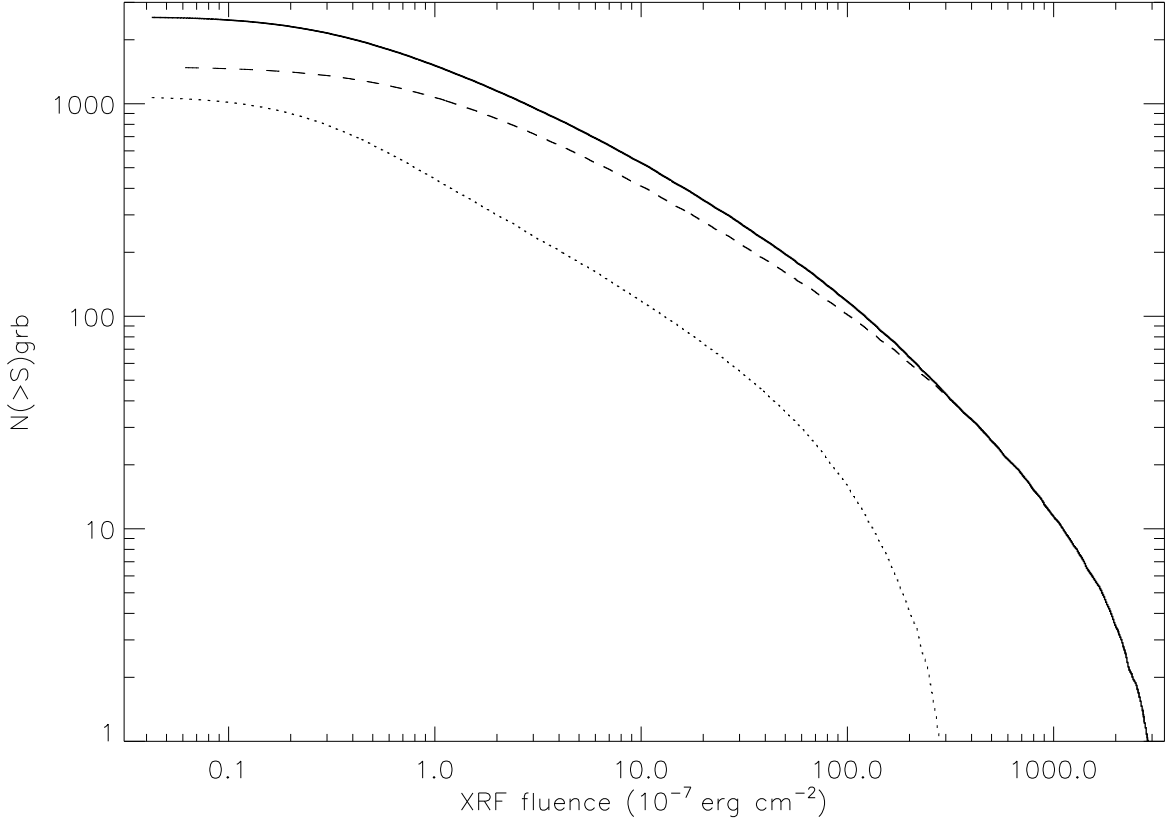}
\caption{
The dashed curve is the expected XRF log N-log S for triggered BATSE GRB,
with $T_{90} > 2$ sec, using the XDGRB estimate of $R_{X/\gamma}$ (as
for the solid line of Fig.4). The dotted curve is the expected contribution 
from untriggered BATSE GRB, assuming 
the same $R_{X/\gamma}$ distribution, and the  solid curve is sum of the 
triggered and untriggered contributions. The small difference suggests that
undetected GRBs are unlikely to make a significant contribution to the FXT
distribution. 
}
\label{trig_untrig}
\end{figure}

\begin{figure}
\plotone{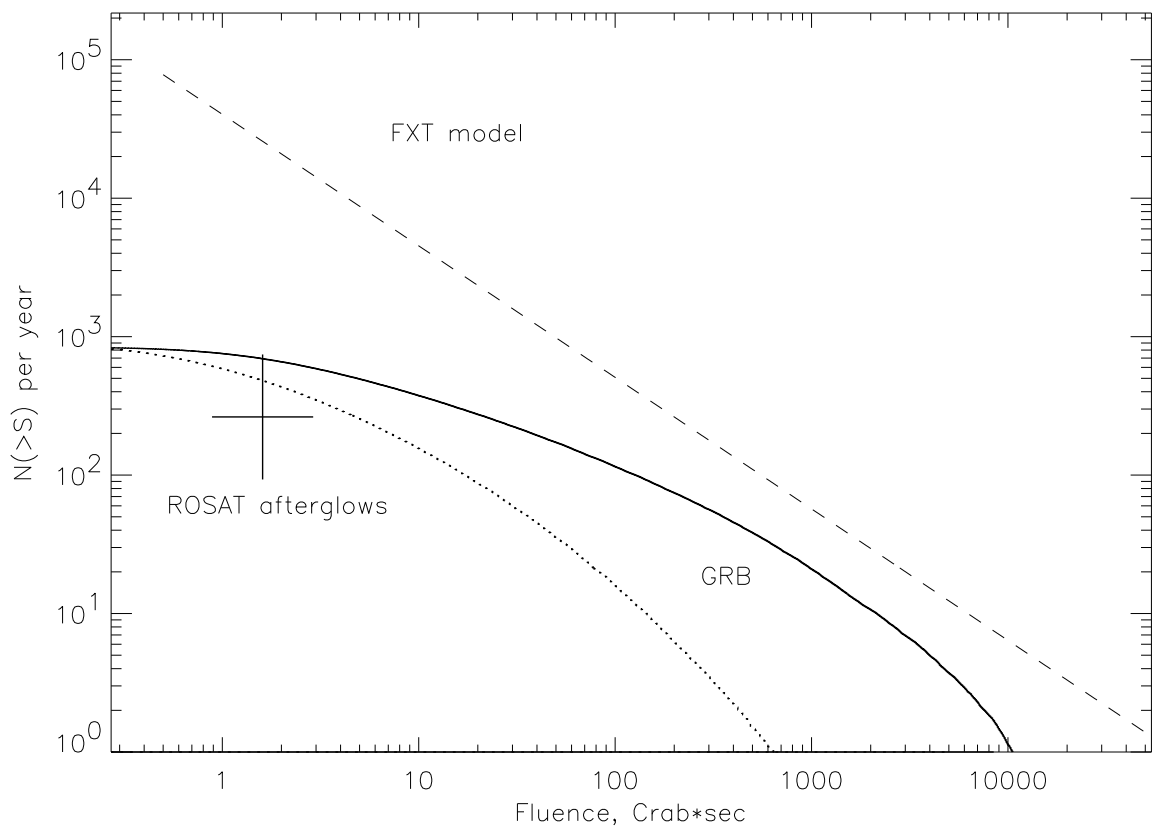}
\caption{Contribution of gamma-ray bursts to the model log N-log S 
distribution for X-ray flashes.
The XRF distribution parameters are taken from Fig.1. The thin solid curve is
the GRB-derived distribution from Fig.6 (XDGRB case), while 
the thin dotted curve is the expected distribution from the `BATSE' case. 
The cross is derived from the ROSAT survey of afterglow-like events
that were not attributed to late-type stars 
\cite{grein00}. Despite the uncertainties in our analysis, it is clear
the classical gamma-ray bursts are more important contributor to bright XRFs than to  faint XRFs.
}
\label{logns_and_grb}
\end{figure}

\begin{figure}
\plotone{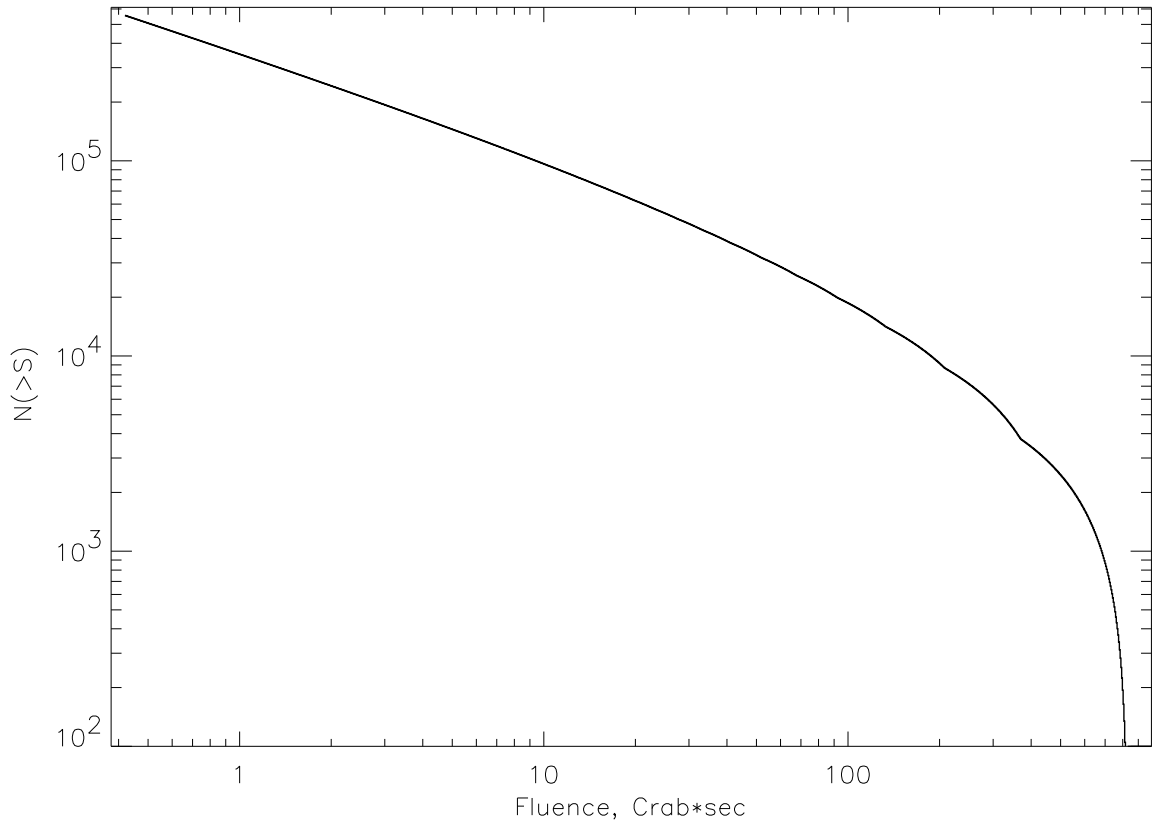}
\caption{The expected log N-log S FXT distribution for late-type
flare stars
in the vicinity of the Sun. The normalization is arbitrary. The
slope for lower fluences is dominated by the intrinsic number-energy distribution of
stellar flares, rather than the 3/2 slope of standard candles in an 
isotropic geometry.
}
\label{Flare_stars}
\end{figure}

\clearpage

\begin{table}
\caption{log($R_{X/\gamma}$) distribution parameters for different models}
\label{tab_1}
\begin{tabular}{ccccccccccc}
\tableline
\tableline
Case & \multicolumn{3}{c}{Distribution} & \multicolumn{7}{c}{Model parameters}\\
 & \multicolumn{3}{c}{parameters} & \multicolumn{2}{c}{Gaussian} & \multicolumn{5}{c}{Gaussian + Quadratic polynomial}\\
 & Mean & Median & $\sigma$ & $\mu$ & $\sigma$ & $\mu$ & $\sigma$ & A$^{c}$ & B$^{c}$ & C$^{c}$\\ 
\tableline
BATSE$^{a}$ & -1.67 & -1.68 & 0.48 & -1.53 & 0.43 & -1.53 & 0.43 & 0.0 & 0.0 & 0.0 \\
XDGRB$^{b}$ & -0.95 & -1.12 & 0.68 & -1.06 & 0.30 & -1.08 & 0.22 & -0.01 & -0.01 & 0.04\\
\tableline
\end{tabular}
\par
\begin{list}{}{}
{\item $^{a}$ BATSE events}
{\item $^{b}$ Ginga + WATCH + RXTE + BeppoSAX events, XDGRB case}
{\item $^{c}$ A, B \& C represent parameters of the quadratic polynomial
$Ax^2+Bx+C$}
\end{list}
\end{table}

\end{document}